\documentclass[12pt,fleqn]{iopart}

\expandafter\let\csname equation*\endcsname=\relax
\expandafter\let\csname endequation*\endcsname=\relax
\usepackage{amsmath}

\usepackage{iopams}
\usepackage{citesort}
\usepackage{epsfig}
\usepackage[caption=false]{subfig}
\usepackage[amssymb,Gray]{SIunits}
\newcommand{\mom}{Q}
\newcommand{\D}{\rmd}

\begin{document}
\ifx\href\undefined\else\hypersetup{linktocpage=true}\fi

\title{Gravity and the Collapse of the Wave Function: 
a Probe into Di\'osi-Penrose model}

\author{Mohammad Bahrami, Andrea Smirne, and Angelo Bassi}

\address{Department of Physics, University of Trieste, 34151 Miramare-Trieste, Italy}

\address{Istituto Nazionale di Fisica Nucleare, Sezione di Trieste, Via Valerio 2, 34127 Trieste, Italy}

\eads{\mailto{mohammad.bahrami@ts.infn.it}, \mailto{andrea.smirne@ts.infn.it}, \mailto{bassi@ts.infn.it}}

\begin{abstract}
We investigate the Di\'osi-Penrose (DP) proposal for connecting the collapse of the wave function to gravity. The DP model needs a free parameter, acting as a cut-off to regularize the dynamics, and the predictions of the model highly depend on the value of this cut-off. The Compton wavelength of a nucleon seems to be the most reasonable cut-off value since it justifies the non-relativistic approach. However, with this value, the DP model predicts an unrealistic high rate of energy increase. Thus, one either is forced to choose a much larger cut-off, which is not physically justified and totally arbitrary, or one needs to include dissipative effects in order to tame the energy increase.
Taking the analogy with dissipative collisional decoherence seriously, we develop a dissipative generalization of the DP model. We show that even with dissipative effects, the DP model contradicts known physical facts, unless either the cut-off is kept artificially large, or one limits the applicability of the model to massive systems. We also provide an estimation for the mass range of this applicability.
\end{abstract}

\section{Introduction} 
By accepting the universality of the quantum superposition principle, it should be possible to observe classical macro objects in a superposition of two distinguishable positions. However, so far no signature of superposition states has been observed at the macro-scale. Macro objects behave classically, while tremendous manifestations of superposition states have been observed at the micro-scale. This state of affairs immediately raises the following questions~\cite{new_phys0,new_phys1,new_phys2,new_phys3}: 
does the superposition principle really hold true at the macro-scale? Is there a division between the micro and the macro world? If so, what is responsible for it?

Collapse models~\cite{Pe1,GRW,d,Pe2,Ghirardi1990b,Cslmass,p,adlerphoto,Bassi2003,a,d1,dd2,Bassi2013} provide a well-defined phenomenology to answer these questions. Initiated by the seminal works of Ghirardi, Rimini, Weber, and Pearle, collapse models assume a universal stochastic noise that couples non-linearly with matter. This non-linear coupling induces a localization in space, which kills superpositions with the correct quantum probabilities. The strength of the coupling is fixed by phenomenological parameters defining the models. The collapse rate then grows by increasing the size and complexity of the system such that the effect of the collapse process is negligible at the micro-scale, and becomes dominant when moving toward the macro-scale. In this way, within a unique dynamical equation, both the quantum and the classical world can be described consistently. 


Collapse models are phenomenological models. Their justification from fundamental physical principles is not yet known and it very much depends on one's view about the physical origin of the collapse field. A natural explanation can be provided by {\it gravity}, because gravity is universal and its strength increases with the mass of the system. In fact, these are two crucial properties of the collapse field.

The connection of the collapse field with gravity has been explored by many authors~\cite{gravity_models1,gravity_models2,gravity_models3,SN,d,dd2,d1,p}, in particular by K\'arolyh\'azy {\it et. al}~\cite{gravity_models2}, Di\'osi~\cite{d,dd2,d1} and Penrose~\cite{p}, independently. 
Here, we will focus on the works of Di\'osi and Penrose, which is usually called as DP model. Di\'osi proposed a stochastic nonlinear Schr\"odinger equation as follows ~\cite{dd2}:
\begin{align}
\label{eq:sch}
\D|\psi_t\rangle&=
\left[-\frac{i}{\hbar}\hat{H}\,\D t
+\int \, \D^3\mathbf{x}\,
\left(\hat{M}({\bf x})-\langle \hat{M}({\bf x})\rangle_t\right)\,\D W_t(\mathbf{x})
\right.\\\nonumber&~~~~~\left.-\frac12
\iint\, \D^3{\bf x}\,\D^3{\bf y}\,{\cal G}({\bf x}-{\bf y})
\left(\hat{M}({\bf x})-\langle \hat{M}({\bf x}) \rangle_t \right)
\left(\hat{M}({\bf y})-\langle \hat{M}({\bf y}) \rangle_t \right)\D t
\right]|\psi_t\rangle,
\end{align}
where $\hat{H}$ is the standard quantum Hamiltonian, 
$\langle \hat{M}({\bf x})\rangle_t=\langle\psi_t| \hat{M}({\bf x})|\psi_t\rangle$, 
$\hat{M}(\mathbf{x})$ is the local mass density operator, which in the first-quantization 
formalism reads: 
\begin{equation}\label{eq:mm}
\hat{M}(\mathbf{x})=\sum_{j=1}^N\,m_j\,\delta({\bf x}-\hat{\bf r}_j),
\end{equation}
with $\hat{\bf r}_j$ the position operator of $j$-th particle; 
and $W_t(\mathbf{x})$ is a real Wiener process producing the white noise $w(t,{\bf x})=\D W_t(\mathbf{x})/\D t$ with the statistical properties:
\begin{eqnarray}
\mathbb{E}(w(t,{\bf x}))&=&0,\qquad
\mathbb{E}(w(t_1,\mathbf{x})\,w(t_2,\mathbf{y}))=
\delta(t_1-t_2)\,{\cal G}({\bf x}-{\bf y}) \label{eq:ww1}
\end{eqnarray}
where $\mathbb{E}(\cdots)$ is the stochastic average, and ${\cal G}({\bf x}-{\bf y})$ the two-point correlation function of the collapse field:
\begin{equation}
{\cal G}({\bf x})= \frac{G}{\hbar}\,\frac{1}{|\mathbf{x}|}, \label{eq:ww2}
\end{equation}
where $G$ is the gravitational constant. With Eq.\eqref{eq:sch} in hand, the statistical operator evolves as 
\begin{equation}
\label{eq:me}
\frac{\partial}{\partial t}\hat{\rho}(t)=
-\frac{i}{\hbar}[\hat{H},\hat{\rho}(t)]
+\mathcal{L}[\hat{\rho}(t)]
\end{equation}
 with:
\begin{eqnarray}
\label{eq:lnd0}
\mathcal{L}[\hat{\rho}(t)]&=&\frac{G}{\hbar}\,\iint\,
\frac{\D^3\mathbf{x}\,\D^3\mathbf{y}}{|\mathbf{x}-\mathbf{y}|}\,\left(
\hat{M}(\mathbf{x})\,\hat{\rho}(t)\,\hat{M}(\mathbf{y})
-\frac12\,\left\{\hat{M}(\mathbf{x})\,\hat{M}(\mathbf{y}),\hat{\rho}(t)\right\}
\right).
\end{eqnarray}

%
However, there are divergent terms in above equation (see Eqs.(\ref{eq:den_many0},\ref{eq:L})). 
To regularize the dynamics, Di\'osi proposed to introduce a cut-off. 
Although the introduction of the cut-off prevents
the divergence in the evolution equation of the statistical operator, 
the energy of the system increases monotonically and goes to infinity. Even more, 
with the originally proposed value of the cut-off, the rate of the energy increase is too high, 
giving rise to the problem of overheating, as shown in~\cite{Ghirardi1990}. 
As a consequence Ghirardi \emph{et al.}~\cite{Ghirardi1990} proposed a much larger cut-off, which, however, is less justified on a physical ground.
We will come back on this issue in section~\ref{sec:overheating}.
An immediate question would be if there is any other resolution for the overheating and the energy divergence problems in the DP model. Here we shall elaborate this question in detail.


 
First, we will show that the model is structurally equivalent to the
the mass-proportional Continuous Spontaneous Localization (CSL) model~\cite{Ghirardi1990,Cslmass}. We will then discuss the connection of DP model with gravity. We will also discuss the analogy of the DP equation to the typical master equations for collisional decoherence. Using this analogy, we will elaborate on the problem of overheating and investigate the possible resolutions. 
The overheating problem should be resolved by introducing dissipative terms, while preserving the translation-covariance of the dynamics. 
However, we will argue that this calls for either a change in the spatial cut-off proposed by Di\'osi or a limitation in the applicability of the model.

\section{Spatial cut-off in the DP model}
To clarify the origin of the divergence in the DP dynamics, we first rewrite Eq.\eqref{eq:lnd0} 
in the form of a diagonal Lindblad master equation \cite{Lindblad1976,Breuer2002}. 
By introducing the inverse Fourier transform of the term $1/|\mathbf{x}-\mathbf{y}|$, one finds:
\begin{eqnarray}
\label{eq:den_many0}
\mathcal{L}[\hat{\rho}(t)]&=&
\frac{G}{2\pi^2\hbar^2}\,\sum_{j,l=1}^Nm_j\,m_l\,
\int\,\frac{\D^3\mathbf{\mom}}{\mom^2}\,
\left(
e^{\frac{i}{\hbar}\,\mathbf{\mom}\cdot\hat{\bf r}_j}\,\hat{\rho}(t)\,
e^{-\frac{i}{\hbar}\,\mathbf{\mom}\cdot\hat{\bf r}_l}
-\frac12\left\{e^{\frac{i}{\hbar}\,\mathbf{\mom}\cdot\hat{\bf r}_j}e^{-\frac{i}{\hbar}\,\mathbf{\mom}\cdot\hat{\bf r}_l} , \hat{\rho}(t)\right\}\right) \nonumber\\
&=&
-\Lambda\,\hat{\rho}(t)+
\frac{G}{2\pi^2\hbar^2}\,\sum_{j=1}^Nm^2_j
\int\,\frac{\D^3\mathbf{\mom}}{\mom^2}\,
e^{\frac{i}{\hbar}\,\mathbf{\mom}\cdot\hat{\bf r}_j}\,\hat{\rho}(t)\,
e^{-\frac{i}{\hbar}\,\mathbf{\mom}\cdot\hat{\bf r}_j} \\\nonumber
&& + \frac{G}{2\pi^2\hbar^2}\,\sum_{j\neq l=1}^Nm_j\,m_l\,
\int\,\frac{\D^3\mathbf{\mom}}{\mom^2}\,
\left(
e^{\frac{i}{\hbar}\,\mathbf{\mom}\cdot\hat{\bf r}_j}\,\hat{\rho}(t)\,
e^{-\frac{i}{\hbar}\,\mathbf{\mom}\cdot\hat{\bf r}_l}
-\left\{e^{\frac{i}{\hbar}\,\mathbf{\mom}\cdot\hat{\bf r}_j}e^{-\frac{i}{\hbar}\,\mathbf{\mom}\cdot\hat{\bf r}_l} , \hat{\rho}(t)\right\}\right),
\end{eqnarray}
where
\begin{equation}
\label{eq:L}
\Lambda=\frac{G}{2\pi^2\hbar^2}\,\sum_{j,l=1}^Nm_j\,m_l\,\int\,
\frac{\D^3\mathbf{\mom}}{\mom^2}
=\frac{2GM^2}{\pi\hbar^2}\,\int_0^{\infty}\D\mom
\end{equation}
with $M=\sum m_j$ the total mass. It is quite clear that $\Lambda$ diverges. To solve this problem, Di\'osi proposed to introduce a cut-off, in order to regularize the dynamics. 

The cut-off can be introduced at the level of Eq.\eqref{eq:lnd0}, by replacing the point-like density operator with a coarse-grained mass density operator, with a spatial resolution $R_0$. Di\'osi originally introduced the coarse-grained mass density as follows:
\begin{equation}\label{eq:dsp}
\hat{M}'({\bf x}) = \frac{3}{4 \pi R^3_0} \int\,\D^3{\bf y}\,
 \theta(R_0-|{\bf x}-{\bf y}|)
\,\hat{M}({\bf y}),
\end{equation}
where $\theta(x)$ is the Heaviside step function. Subsequently, 
Ghirardi {\it et al.}~\cite{Ghirardi1990} introduced the coarse-graining as follows:
\begin{equation}\label{eq:co}
\hat{M}'({\bf x})=(2\pi R_0^2)^{-3/2}\int\,\D^3{\bf y}\,
\exp\left(-\frac{|{\bf x}-{\bf y}|^2}{2R_0^2}\right)\,\hat{M}({\bf y}).
\end{equation}
Note that $\hat{M}'({\bf x})$ is meant to replace $\hat{M}({\bf x})$ in Eq.\eqref{eq:lnd0}.
Accordingly, Eq.\eqref{eq:den_many0} becomes:
\begin{eqnarray}
\nonumber
\mathcal{L}[\hat{\rho}(t)] &=&
\frac{G}{2\pi^2\hbar^2}\,\sum_{j,l=1}^Nm_j\,m_l\,
\int\,\frac{\D^3\mathbf{\mom}}{\mom^2}\,
f(\mom)\,\times
\\&&\qquad\qquad
\label{eq:den_many}
\left(
e^{\frac{i}{\hbar}\,\mathbf{\mom}\cdot\hat{\bf r}_j}\,\hat{\rho}(t)\,
e^{-\frac{i}{\hbar}\,\mathbf{\mom}\cdot\hat{\bf r}_l}
-\frac12\left\{e^{\frac{i}{\hbar}\,\mathbf{\mom}\cdot\hat{\bf r}_j}e^{-\frac{i}{\hbar}\,\mathbf{\mom}\cdot\hat{\bf r}_l} , \hat{\rho}(t)\right\}\right),
\end{eqnarray}
where $f(\mom)$ is a damping function of the momentum $\mom$. For the coarse-graining in Eq.\eqref{eq:dsp} we have:
\begin{equation}
f(\mom) = \frac{9 \hbar^6}{R^6_0 \mom^6}\left(\sin\left(\frac{\mom R_0}{\hbar}\right) -\frac{\mom R_0}{\hbar} \cos\left(\frac{\mom R_0}{\hbar}\right) \right)^2,
\end{equation}
while for the coarse-graining as in Eq.\eqref{eq:co}, we find:
\begin{equation}
\label{eq:gua}
f(\mom) = \exp\left(-\frac{\mom^2 R_0^2}{\hbar^2}\right),
\end{equation}
However, since $e^{-x^2/2} \approx 3(\sin x - x \cos x)/x^3$,
the two different ways of introducing the cut-off are practically equivalent,
once $R_0$ is fixed. For the sake of simplicity, we will use the coarse-graining given in Eq.\eqref{eq:co}. 


The cut-off can be also included at the level of Eq.\eqref{eq:den_many0} as follows:
\begin{eqnarray}
\mathcal{L}[\hat{\rho}(t)]&=&
\frac{G}{2\pi^2\hbar^2}\,\sum_{j,l=1}^Nm_j\,m_l\,
\int_0^{\mom_{\text{\tiny max}}}\D\mom \iint \D^2\tilde{\bf n}\,\times\nonumber\\&&
\label{eq:den_many3}
\qquad\qquad\left(
e^{\frac{i}{\hbar}\,\mom\tilde{\bf n}\cdot\hat{\bf r}_j}\,\hat{\rho}(t)\,
e^{-\frac{i}{\hbar}\,\mom\tilde{\bf n}\cdot\hat{\bf r}_l}
-\frac12\left\{e^{\frac{i}{\hbar}\,\mom\tilde{\bf n}\cdot\hat{\bf r}_j}e^{-\frac{i}{\hbar}\,\mom\tilde{\bf n}\cdot\hat{\bf r}_l} , \hat{\rho}(t)\right\}\right),
\end{eqnarray}
where $\mom_{\text{\tiny max}}=\hbar/R_0$.
Here, $\tilde{\bf n}={\bf Q}/Q$, and $\D^2\tilde{\bf n}=\D\cos\theta\,\D\phi$ is the
corresponding solid angle differential~\footnote{It is quite clear that Eq.\eqref{eq:den_many3} is mathematically equivalent to Eq.\eqref{eq:den_many} with 
$f(\mom)=\theta(\mom-\mom_{\text{\tiny max}})$, which eventually can be approximated by a damping Gaussian function, $f(\mom)\approx \exp(-\mom^2/\mom^2_{\text{\tiny max}})$, such as Eq.\eqref{eq:gua}.}. 
At the level of Eq.\eqref{eq:den_many3}, we can assign an interpretation to the cut-off and even justify a specific value for it. In fact, $\mom_{\text{\tiny max}}$ can be read as an upper limit for the modes of the collapse field that are the dominant modes contributing to the collapse. These modes are also small enough to justify the non-relativistic approach. This interpretation is very similar to Bethe's non-relativistic computation of the Lamb shift (e.g., see~\cite{mw}). Thus, if we argue that $\mom_{\text{\tiny max}}$ is the bound justifying the non-relativistic approach, then one can replace $R_0$ by the Compton wavelength; that is to say: $R_0=\frac{2\pi\hbar}{mc}$, which is $R_0\approx10^{-15}\,$m for a nucleon. Accordingly, in this way, one can provide a justification for the cut-off and its chosen value. This value corresponds to the original choice made by Di\'osi~\cite{d}; which however is incompatible with experimental data. We will discuss this issue in larger detail in section~\ref{sec:overheating}.

\section{Physical Meaning and Solutions of the DP Master Equation} 
In this section, we will focus
on the regularized one-particle DP model where the dynamics reads:
\begin{equation}
\label{eq:den}
\mathcal{L}[\hat{\rho}(t)]=
\int\,\D^3\mathbf{\mom}\,\, \Gamma_{\text{\tiny DP}}({\bf \mom})\,
\left(
e^{\frac{i}{\hbar}\,\mathbf{\mom}\cdot\hat{\mathbf{r}}}\,\hat{\rho}(t)\,
e^{-\frac{i}{\hbar}\,\mathbf{\mom}\cdot\hat{\mathbf{r}}}
-\hat{\rho}(t)\right),
\end{equation}
with 
\begin{equation}
\label{rate}
\Gamma_{\text{\tiny DP}}({\bf \mom}) = \frac{G m^2}{2\pi^2\hbar^2}\,\frac{1}{\mom^2}\,
\exp\left(-\frac{\mom^2R_0^2}{\hbar^2}\right).
\end{equation} 

The master equation given in Eq.\eqref{eq:den} is more convenient than the corresponding one in Eq.\eqref{eq:lnd0} for several reasons.
First, Eq.\eqref{eq:den}
directly shows that we have a completely positive and translation-covariant
semigroup dynamics~\cite{Holevo1993}. In addition, it allows to deduce a direct analogy with models for collisional decoherence. 
The Lindblad structure in Eq. \eqref{eq:den} is in fact the same as that which characterizes collisional decoherence of a very massive particle interacting through collisions with a low density background gas~\cite{gf,a1}. 
Eq.\eqref{eq:den} describes a pure decoherence dynamics with no dissipation~\cite{Vacchini2009}. In particular,
the total transition rate, i.e. the probability per unit time
that the particle undergoes a collision, is 
\begin{eqnarray}
\label{eq:rate_0}
\Lambda_{\text{\tiny DP}}=\int\,\D^3{\bf \mom}\,\,\Gamma_{\text{\tiny DP}}({\bf \mom})=\frac{Gm^2}{\sqrt{\pi}\hbar\,R_0},
&&\text{with the dimension }\,
\left[\Lambda_{\text{\tiny DP}}\right]=\text{s}^{-1}.
\end{eqnarray}

With Eq.\eqref{eq:den} in hand, one can also easily find the solution of the DP master equation,
at least for a single free particle.
Using the characteristic function~\cite{Savage1985,Smirne2010}, the state at time $t$ in the position representation, $\langle{\bf x}|\hat{\rho}(t)|\mathbf{x}'\rangle = \rho(\mathbf{x}, \mathbf{x}',t)$, is found to be
\begin{eqnarray}
\label{eq:sol}
\rho(\mathbf{x}, \mathbf{x}', t)  &=&  \iint
\D^3\mathbf{y}\,\frac{\D^3 {\bf p}}{(2 \pi \hbar)^3} 
\,\rho_0 (\mathbf{x}+\mathbf{y}, \mathbf{x}'+\mathbf{y}, t)\times\nonumber
\\&&\qquad
\exp\left(-\frac{i}{\hbar}{\bf y} \cdot \mathbf{p}
- \frac{1}{\hbar}\int^t_0\,d\tau\left(
U \left(-\frac{{\bf p}\,\tau}{ m} +\mathbf{x}-\mathbf{x}'\right)
-U(0)\right)\right),
\label{eq:sol}
\end{eqnarray}
where $ \rho_0(\mathbf{x}, \mathbf{x}',t)$ is the solution of the free Schr{\"o}dinger dynamics and
\begin{equation}
U(\mathbf{x})= - G\,\iint \frac{\D^3 \mathbf{r} \,\D^3 \mathbf{r}'   M'(\mathbf{r}) \,M'(\mathbf{r}')}
{|\mathbf{x}+\mathbf{r}-\mathbf{r}'|}
=
-G m^2\,\frac{\text{Erf}(|{\bf x}|/2R_0)}
{|{\bf x}|},
\end{equation}
is the Newtonian self-interaction where Erf$(x)$ is the error function. 
If one neglects the pure Schr\"odinger contribution in Eq.\eqref{eq:me}, which is justified on the short time scale, then Eq.\eqref{eq:sol} reduces to an exponential
decay of the form: 
\begin{equation}
\label{eq:damp}
\rho(\mathbf{x}, \mathbf{x}', t) = \exp\left(- \frac{t}{\tau(\mathbf{x}, \mathbf{x}')}\right)\rho(\mathbf{x}, \mathbf{x}', 0),
\end{equation}
where the characteristic damping time $\tau$ is:
\begin{equation}
\label{eq:td}
\tau(\mathbf{x}, \mathbf{x}') = \frac{\hbar}
{U \left(\mathbf{x}-\mathbf{x}'\right)-U(0)}.
\end{equation}
For $|{\bf x}-{\bf x'}| \gg R_0$, one finds that
$\tau({\bf x},{\bf x'}) \approx \Lambda^{-1}_{\text{\tiny DP}}$, which is of the order $\tau \approx 10^{15}\,\text{s}$ with $m=1\,$amu and $R_0=10^{-15}\,$m.
As usual in collapse models \cite{Bassi2013}, we will thus exploit $\Lambda_{\text{\tiny DP}}$ to estimate the collapse rate
of the model.

Equation \eqref{eq:damp} implies that spatial superpositions of positions ${\bf x}$ and ${\bf x'}$ decays with the rate $\tau({\bf x},{\bf x}')$. 
This is precisely Penrose's idea~\cite{Penrose1996,gao,Bassi2013}. According to Penrose, a spatial superposition of matter generates the superposition of two bumps in the space-time. In this situation, the time-translation operator is ill-defined. This ill-definedness leads to an uncertainty in the energy which is given, in the weak-field limit, by the Newtonian gravitation self-energy of the corresponding mass distribution at two positions, i.e. $\Delta E=U \left(\mathbf{x}-\mathbf{x}'\right)-U(0)$. Using the energy-time uncertainty relation, this can imply that the superposition decays to one of the localized states with the lifetime $\tau=\hbar/\Delta E$, which is equivalent to Eq.\eqref{eq:td}. Therefore, Di\'osi's dynamical equation implements Penrose's idea. This is why one speaks of Di\'osi-Penrose model.

However, it should be pointed out that the DP model is not a gravity-induced collapse model, as was clear to the original proposers who always spoke in terms of gravity-{\it related} models~\cite{d-private}. 
The DP dynamics in Eq.\eqref{eq:sch} is simply postulated and there is no real physical justification for it besides the requirements of the state vector normalization and no faster than light signaling~\footnote{This is true for all collapse models. For the DP model, Di\'osi provided a partial justification for the existence of a universal Newtonian noise. Following Bohr-Rosenfeld argument on uncertainties of measurement of the electromagnetic fields, Di\'osi and Luk\'acs introduced a universal bound for the optimal measurement of the gravitational acceleration field (see~\cite{d} and references therein). Then, they interpreted this bound as representation of a universal fluctuation in the gravitational field, which is represented as a Newtonian noise. Nevertheless, the proposed form of the dynamic (Eq.\eqref{eq:sch}) and how the noise couples with the quantum system is ultimately still postulated, not derived}. In addition, gravity only appears in the correlation function of the collapse field as a Newtonian potential; there is no gravitational interaction between matter and gravity in the DP model.
Therefore, Penrose's original idea still remains without a satisfactory explanation, not even at the phenomenological level.

\section{Comparison of the DP Model with the CSL Model} 
As pointed out in \cite{Ghirardi1990}, the DP collapse model shares strong similarities with
the mass-proportional Continuous Spontaneous Localization (CSL) model~\cite{Ghirardi1990,Cslmass}. 
The CSL model is originally formulated in terms of quantum field theory~\cite{Ghirardi1990b,Bassi2013}. 
However, if we restrict to the $N$-particle sector of the Fock space, its defining stochastic differential equation
can be expressed through the local mass density operator 
$\hat{M}({\bf x})$ as given in Eq.\eqref{eq:mm}.
Explicitly, we have
\begin{eqnarray}
\label{eq:csl}
\frac{\partial}{\partial t} |\psi_t\rangle&=& \left[- \frac{i}{\hbar} \hat{H} + \frac{\sqrt{\gamma}}{m_0} \int \D^3 {\bf x}\,\left(\hat{M}({\bf x})-\langle \hat{M}({\bf x}) \rangle_t \right)\, 
\D W_t({\bf x})\right.\\\nonumber
&&\left.
- \frac{\gamma}{2m^2_0} \iint \D^3{\bf x}\,\D^3{\bf y}\,{\cal G}({\bf x}-{\bf y})
\left(\hat{M}({\bf x})-\langle \hat{M}({\bf x}) \rangle_t \right)
\left(\hat{M}({\bf y})-\langle \hat{M}({\bf y}) \rangle_t \right)  \right] |\psi_t\rangle,
\end{eqnarray}
where
$m_0=1\,$amu is a reference mass, and $\gamma=10^{-36}\,\text{m}^3\cdot\text{s}^{-1}$ is a new intrinsic parameter of the model setting the collapse strength~\cite{GRW,Ghirardi1990b,Cslmass};
note that a much higher value, $\gamma=10^{-28\pm2}\,\text{m}^3\cdot\text{s}^{-1}$, was proposed more recently~\cite{adlerphoto}.
In addition, $\mathcal{G}({\bf x})$ is a Gaussian function:
\begin{equation}\label{eq:ggx}
\mathcal{G}({\bf x}) = \frac{1}{(4 \pi r_c)^{3/2}} \exp\left(-\frac{{\bf x}^2}{4 r^2_c}\right),
\end{equation}
where $r_c = 10^{-7} \text{m}$ is the second new intrinsic parameter of the model.
The white-noise field $w(t,{\bf x}) = d W_t({\bf x})/ d t$ satisfies Eq. \eqref{eq:ww1}
and Eq.\eqref{eq:ggx} replaces Eq.\eqref{eq:ww2}.

Manifestly, the structure of the CSL equation in Eq.\eqref{eq:csl} is 
the same as that of the DP dynamics given in Eq.\eqref{eq:sch}. 
The only difference is the spatial correlation function of the noise: a Gaussian function in the CSL model and the Newtonian potential in the DP model.
The DP choice, which is meant to establish a connection with gravity, requires the introduction of a spatial cut-off in order to avoid divergences, as we discussed before. 
The CSL model on the other hand requires two new parameters $\gamma$ and $r_c$ where $r_c$ defines the correlation function of the noise, and $\gamma$ sets the strength of the localization processes. In the DP model the strength of the localization processes is set by the constant $G$, as a phenomenological signature of the gravity.
However, because of the divergence problem, one needs to introduce a new free parameter $R_0$,
which plays a role analogous to $r_c$. We shall elaborate this issue by looking at the CSL and DP master equations.

For a one particle system the CSL master equation is equivalent to that of the collapse model introduced by Ghirardi, Rimini and Weber (GRW model)~\cite{GRW}
and it reads
\begin{eqnarray}\label{grw}
\frac{\partial}{\partial t}\hat{\rho}(t)&=&
-\frac{i}{\hbar}\left[\hat{H},\hat{\rho}(t)\right] + \int\, \D^3 \mathbf{\mom} \, \, \Gamma_{\mbox{\footnotesize{{CSL}}}} (\mathbf{\mom}) 
\left(
e^{\frac{i}{\hbar}\,\mathbf{\mom}\cdot\hat{\mathbf{r}}}\,\hat{\rho}(t)\,
e^{-\frac{i}{\hbar}\,\mathbf{\mom}\cdot\hat{\mathbf{r}}}
-\hat{\rho}(t)\right),
\end{eqnarray}
where
\begin{equation}\label{rategrw}
\Gamma_{\mbox{\footnotesize{{CSL}}}} (\mathbf{\mom}) =
\frac{\gamma}{(2 \pi \hbar)^{3}} \, \frac{m^2}{m^2_0}\,
\exp\left(- \frac{\mom^2 r^2_c}{ \hbar^2}\right);
\qquad m_0=1\text{amu}.
\end{equation}
The difference with the DP master equation given in Eq.\eqref{eq:den} is entirely contained in the rate term (compare with Eq.\eqref{rate}).
Apart from the different coefficients in front of the Gaussian distribution,
the main differences are the factor $1/\mom^2$, which is present only in the case of the DP model,
and, crucially, the width of the Gaussian, which is set by $1/r_c$ in the CSL model and 
by $1/R_0$ in the DP model.
It means that the momentum fluctuations are indeed much larger in the DP model,
giving rise to an unacceptable rate of the energy increase. We shall discuss this issue later.

The solution of the master equation associated with the CSL model has the same
form as in Eq.\eqref{eq:sol} where one has to replace $U({\bf x})$ by $-\hbar\Phi({\bf x})$ where:
\begin{equation}
\Phi({\bf x}) = \int \D^3 {\bf \mom}\,\, e^{\frac{i}{\hbar} {\bf \mom} \cdot {\bf x}}\, \Gamma_{\mbox{\footnotesize{{CSL}}}} (\mathbf{\mom}) = \frac{m^2}{m^2_0}\, \frac{\gamma}{(4 \pi r_c)^{3/2}}\, \exp\left(- \frac{x^2}{4 r^2_c}\right).
\end{equation}
Accordingly, like Eq.\eqref{eq:damp}, the CSL decoherence time becomes:
\begin{equation}
\tau_{\text{\tiny CSL}}({\bf x}, {\bf x}') = \frac{1}{\Phi(0) - \Phi({\bf x}-{\bf x}')}.
\end{equation}
For $|{\bf x}-{\bf x}'| \gg r_c$, one has $\tau_{\text{\tiny CSL}}({\bf x}, {\bf x}') \approx \Lambda^{-1}_{\text{\tiny CSL}}$,
where the CSL collapse rate is given by
\begin{eqnarray}
\label{eq:ratecsl}
\Lambda_{\text{\tiny CSL}}=\int\,\D^3{\bf \mom}\,\,\Gamma_{\text{\tiny CSL}}({\bf \mom})= \frac{m^2}{m^2_0} \frac{\gamma}{(4 \pi r^2_c)^{3/2}}.
\end{eqnarray}

\begin{figure}
\hspace{2.5cm}
\includegraphics[width=10cm,keepaspectratio=true]{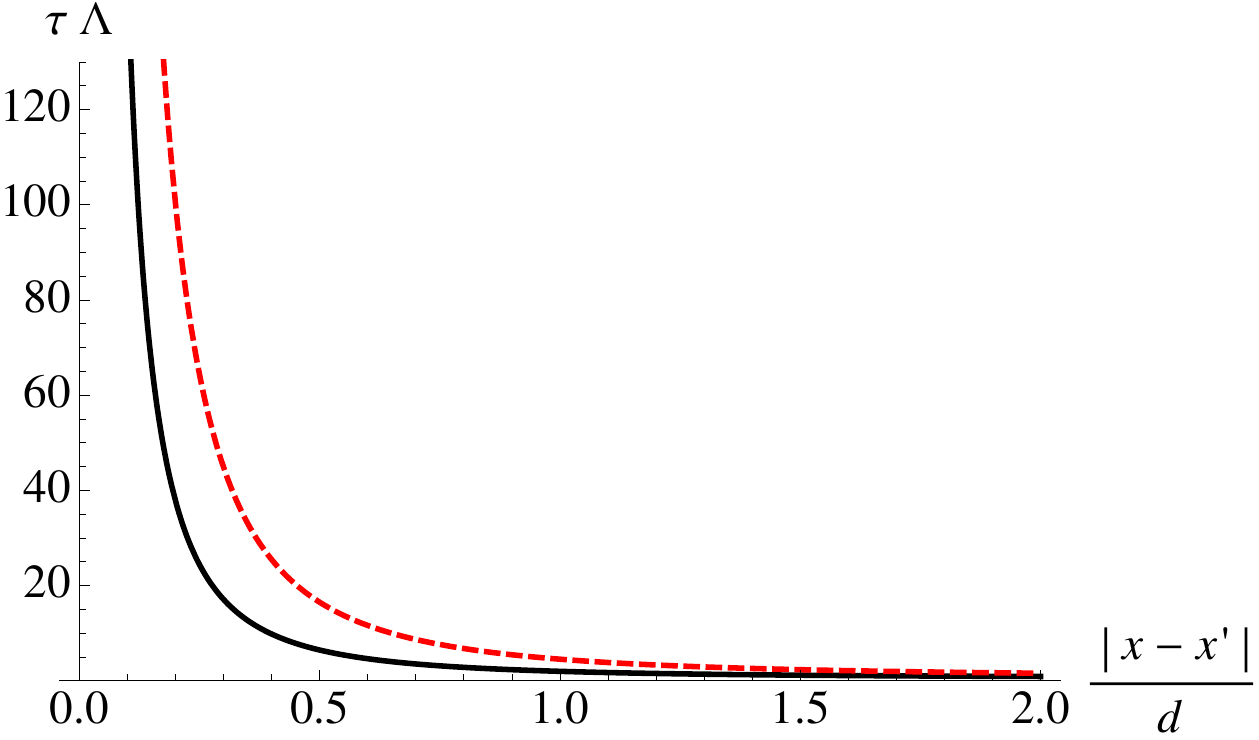}
\caption{Damping time $\tau({\bf x},{\bf x'})$ as a function of the distance $|{\bf x}-{\bf x'}|$ for the DP model (black line) and for the
CSL model (red, dashed line). The damping time is expressed in units
of the inverse of the rate $\Lambda$ for the corresponding model: respectively,
$\Lambda_{\text{\tiny DP}}  \approx 10^{-15}~$s, see Eq.(\ref{eq:rate_0}), and $\Lambda_{\text{\tiny CSL}}\approx 2.2 \times 10^{-17}~$s, see Eq.(\ref{eq:ratecsl}), where both numerical values refer to the case of a nucleon.
The distance is expressed in units of $d$, where, crucially, $d = R_0$ for the DP model and $d=r_c$
for the CSL model.  
}\label{fig:1}
\end{figure}

We plot in Fig.\ref{fig:1} the DP and CSL decoherence times as a function of the distance $|{\bf x}-{\bf x}'|$.
One can see how the damping time decays as a function of the distance in a very 
similar fashion in the two cases. The crucial difference is that the decay length is
fixed by the localization width $r_c$ in the CSL model, while it is fixed by the cut-off $R_0$
in the DP model. As a consequence, the superposition between two states centered
around two different positions at a fixed distance $|{\bf x}-{\bf x'}|$ is suppressed
much more quickly in the DP model.

\section{Amplification Mechanism} 
A key feature of collapse models is amplification mechanism, ensuring the classicality when moving toward 
the macro-scale. It implies that the collapse mechanism suppresses the Schr\"odinger dynamics for the center-of-mass motion of a macro-system. It can be formulated as follows.
Assuming 
a rigid many-body system and tracing out the relative coordinates, the dynamical equation for the center-of-mass takes
the same form as in Eq.\eqref{eq:den}:
\begin{equation}
\label{eq:DP_CM}
{\cal L}[\hat{\rho}_{\text{\tiny M}}(t)]=
\int\,\D^3\mathbf{\mom}\, \Gamma_{\text{\tiny M}}({\bf \mom})
\left(
e^{\frac{i}{\hbar}\,\mathbf{\mom}\cdot\hat{\mathbf{R}}}\,\hat{\rho}_{\text{\tiny M}}(t)\,
e^{-\frac{i}{\hbar}\,\mathbf{\mom}\cdot\hat{\mathbf{R}}}
-\hat{\rho}_{\text{\tiny M}}(t)\right),
\end{equation}
where $\hat{\rho}_{\text{\tiny M}}(t)$ is the center-of-mass density matrix, $\hat{\mathbf{R}}$ is the position operator of the center-of-mass and: 
\begin{eqnarray}
\Gamma^{\text{\tiny M}}_{\text{\tiny DP}}({\bf \mom})&=&
\frac{G}{2\pi^2\hbar^2\mom^2}\,
|\tilde{\varrho}_{\text{rel}}({\bf \mom})|^2\,\exp\left(-\frac{\mom^2R_0^2}{\hbar^2}\right)
\end{eqnarray}
with
$\tilde{\varrho}_{\text{\tiny rel}}({\bf \mom})=\int\,\D^3{\bf x}\,e^{i{\bf \mom}\cdot{\bf x}/\hbar}\,\varrho_{\text{\tiny rel}}({\bf x})$ where $\varrho_{\text{\tiny rel}}({\bf x})$ is the internal mass density. For example, for a homogeneous rigid sphere of mass $M$ and radius $R$, we get: 
$\tilde{\varrho}_{\text{\tiny rel}}({\bf \mom})\approx M\,\exp(-\mom^2R^2/2\hbar^2)$. Accordingly, we find:
\begin{equation}
\Gamma^{\text{\tiny M}}_{\text{\tiny DP}}({\mom})\approx \frac{G\,M^2}{2\pi^2\hbar^2}\,\frac{1}{\mom^2}
\exp\left(-\frac{\mom^2(R^2+R^2_0)}{\hbar^2}\right).
\end{equation}
Similar to Eq.\eqref{eq:rate_0}, here the total collapse rate becomes:
\begin{eqnarray}
\label{eq:rate_M}
\Lambda^{\text{\tiny M}}_{\text{\tiny DP}}&=&\int\,\D^3{\bf \mom}\,\,\Gamma_{\text{\tiny M}}({\bf \mom})\approx
\frac{GM^2}{\,\hbar\,\sqrt{\pi(R^2+R^2_0)}}.
\end{eqnarray}
For example, $\Lambda^{\text{\tiny M}}_{\text{\tiny DP}}$ is of order $10^{-5}\,$s$^{-1}$ for a typical optomechanical nanosphere with $M\approx10^{9}\,$amu and $R\approx50\,$nm~\cite{opto}. Evidently, the collapse rate of a nanosphere is $10$ orders of magnitude larger than of a nucleon, which is of order $10^{-15}\, \text{s}^{-1}$. This is a manifestation of the amplification mechanism.

The center-of-mass master equation for the mass-proportional CSL model is the same as Eq.\eqref{eq:DP_CM}, where:
\begin{equation}
\Gamma_{\text{\tiny CSL}}^{\text{\tiny M}}({\mom})\approx \frac{\gamma}{(2\pi\hbar)^3}\,\frac{M^2}{m_0^2}\,
\exp\left(-\frac{\mom^2(R^2+r_C^2)}{\hbar^2}\right).
\end{equation}
Likewise, one finds:
\begin{eqnarray}
\label{eq:rate_M}
\Lambda_{\text{\tiny CSL}}^{\text{\tiny M}}&=&\int\,\D^3{\bf \mom}\,\,\Gamma_{\text{\tiny M}}({\bf \mom})\approx
\frac{\gamma\,M^2}{8\pi^{3/2}\,m_0^2\,(R^2+r^2_C)^{3/2}}.
\end{eqnarray}
For example, $\Lambda_{\text{\tiny CSL}}^{\text{\tiny M}}$ is of order $10~$s$^{-1}$ for a typical optomechanical nanosphere with $M\approx10^{9}\,$amu and $R\approx50\,$nm~\cite{opto}.

Although the values of the DP and the CSL collapse rates are different, the collapse rate has the same dependence on the mass in both models, and it is proportional to the square of the total mass.

\section{The Overheating Problem}
\label{sec:overheating}
The dynamical equation of the form given in Eq.\eqref{eq:den} does not conserve the energy~\cite{Bassi2005,Vacchini2007}. The rate of energy increase can be easily calculated
and turns out to be:
\begin{eqnarray}\label{eq:hh}
\frac{\D E_{\text{\tiny DP}}(t)}{\D t} &=& \frac{2 \pi}{m} \int^{\infty}_0 \D \mom \,\, \Gamma_{\text{\tiny DP}}(\mom)\, \mom^4 = \frac{m \,G \,\hbar}{4\sqrt{\pi}\,R_0^3},
\end{eqnarray}
with $E_{\text{\tiny DP}}(t)=\text{tr}[\hat{\rho}(t) \hat{H}]$ where $\hat{\rho}(t)$ satisfies the DP dynamics.
From this relation, one can easily  evaluate
the different implications of the cut-off proposed, respectively,
by Di\'osi~\cite{d} and Ghirardi {\it et al.}~\cite{Ghirardi1990}.
In the former case, $R_0 = 10^{-15} \text{m}$,
one gets a rate for the energy increase of order $10^{-4}\,$K/s for a proton, which means a thermal catastrophe!
The second choice, $R_0 = 10^{-7} \text{m}$,
leads to a rate of order $10^{-28}\,$K/s for a proton, which is indeed
a much more reasonable value.
Although in this way the problem of overheating has been partially resolved,
it s clear that the introduction of a cut-off $R_0 = 10^{-7} \text{m}$ is much less justified
than the original proposal by Di{\'o}si. One of the main motivations of the DP model,
i.e., to provide a phenomenological model without free parameters,
is in this way lost.
A crucial question is thus whether it is possible to resolve the overheating problem in a different way.
In the next section, we shall investigate such a possibility.

\section{Dissipative DP dynamics} 
If one takes the DP model---like all collapse models---
seriously, then one is assuming the existence of a classical
random field, filling space,
which couples non-linearly to quantum matter causing the collapse (so far, in a non better specified way,
other than postulating ad hoc equations of motion). Seen from this perspective, there is no surprise
as to why the energy is not conserved: dissipative terms are neglected.
To make a classical analogy, it is like considering a particle in a gas, without
taking dissipation into account. The gas then acts like an infinite temperature gas, increasing
the energy of the particle constantly in time, till it eventually diverges.
The natural resolution to the problem is to include dissipation.
The gas then acts like a real physical gas with a finite temperature $T$, and the particle more
or less quickly thermalizes to that temperature. Clearly,
the resulting equations of motion will be more complicated (uglier, if one wishes),
but will describe a more realistic situation.
More complicated equations of motion are not a problem here,
as we are dealing with phenomenological models.
Now, we will apply this idea to the DP model.

Very similar to collisional decoherence, dissipation can be introduced by replacing $\Gamma_{\text{\tiny DP}}$ in Eq.\eqref{eq:den} 
with an operator-valued function of the system's momentum operator $\hat{\bf p}$. 
To preserve the translation-covariance of the Lindblad structure, one needs a master equation in the form \cite{Holevo1993}:
\begin{equation}
\label{eq:diss}
\mathcal{L}[\hat{\rho}(t)]=\,\int\, \D^3\mathbf{\mom}
\left(
e^{\frac{i}{\hbar}\,\mathbf{\mom}\cdot\hat{\mathbf{r}}}\,
\hat{L}(\mathbf{\mom},\mathbf{\hat{p}}) \hat{\rho}(t) \hat{L}^{\dag}(\mathbf{\mom},\mathbf{\hat{p}})\,
e^{-\frac{i}{\hbar}\,\mathbf{\mom}\cdot\hat{\mathbf{r}}}
-\frac{1}{2} \left\{\hat{L}^{\dag}(\mathbf{\mom},\mathbf{\hat{p}})\hat{L}(\mathbf{\mom},\mathbf{\hat{p}}) \, , \, \hat{\rho}(t)\right\}\right).\;
\end{equation}
In particular, we propose the following choice for $\hat{L}(\mathbf{\mom},\mathbf{\hat{p}})$:
\begin{equation}
\label{eq:S0_Q}
\hat{L}(\mathbf{\mom},\mathbf{\hat{p}}) \equiv  
 \frac{m\sqrt{G}}{\pi\sqrt{2}\,\hbar} 
 \,\frac{1}{\mom} \,
 \exp\left[-\frac{R^2_0}{2\hbar^2}
\left((1+k)\, {\mom}+ 2k\, \frac{\hat{\mathbf{p}}\cdot{{\bf \mom}}}{\mom}\right)^2\right],
\end{equation}
where $k$ is a dimensionless parameter inversely proportional to the mass, given by:
\begin{equation}\label{eq:alm}
k =\frac{m_r}{m},
\end{equation}
with $m_r$ a reference mass.
By setting $m_r=0$ one recovers the DP master equation without dissipation. 
With our specific choice of $k$ in Eq.\eqref{eq:alm}, which eventually appears in $\hat{L}(\mathbf{\mom},\mathbf{\hat{p}})$ too, we will be able to associate the collapse field with a temperature
which does not depend on the mass of the system (see Eq.(\ref{eq:temp})). 
This is a very desirable requirement. 
Of course different choices for 
the operator $\hat{L}(\mathbf{\mom},\mathbf{\hat{p}})$ are possible. 
The one given in Eq.\eqref{eq:S0_Q} 
corresponds to the master equation which describes
the dissipative collisional decoherence in the weak coupling regime~\cite{Vacchini2000,Vacchini2009}:
\begin{equation}
\label{eq:S0_col}
\hat{L}_{\text{\tiny DEC}}(\mathbf{\mom},\mathbf{\hat{p}}) \propto
 \exp\left[-\frac{1}{16m_{\text{\tiny en}}k_BT}
\left((1+\frac{m_{\text{\tiny en}}}{m})\, {\mom}+ 2\,\frac{m_{\text{\tiny en}}}{m}\, \frac{\hat{\mathbf{p}}\cdot{{\bf \mom}}}{\mom}\right)^2\right], 
\end{equation}
with $m_{\text{\tiny en}}$ the mass of environment particles.
Comparing Eq.\eqref{eq:S0_Q} with Eq.\eqref{eq:S0_col}, one can in fact infer the following analogies:
\begin{equation}\label{eq:aT}
m_r\, \longleftrightarrow m_{\text{\tiny en}};\qquad\qquad
R_0 \longleftrightarrow \frac{\hbar}{\sqrt{8 m_{\text{\tiny en}}k_BT}},
\end{equation}
where $\hbar/\sqrt{m_{\text{\tiny en}}k_BT}$ is the thermal de Broglie wavelength of the environment.

\section{The Overheating Problem in Dissipative Mode} 
With Eq.\eqref{eq:diss} in hand, we can show explicitly that
the mean value of the energy is finite also for large times, because of the dissipative effects. The general structure of the master 
equation in Eq.\eqref{eq:diss} implies that the mean value of the kinetic energy $\hat{H}=\hat{p}^2/2m$ satisfies the following
equation:
\begin{eqnarray}
\frac{\D}{\D t} E_{\text{\tiny DP}}(t) &=&\frac{1}{2m}\int \D^3\mathbf{\mom} 
\,\text{Tr}\left(\hat{\rho}(t)\,
 \hat{L}^{\dag}(\mathbf{\mom},\hat{\mathbf{p}})\hat{L}(\mathbf{\mom},\hat{\mathbf{p}})\,(\mom^2 + 2 \, \hat{\mathbf{p}}\cdot \mathbf{\mom}) \right). 
\end{eqnarray}
With the operator $\hat{L}(\mathbf{\mom},\hat{\mathbf{p}})$ as given in Eq.\eqref{eq:S0_Q},
we find:
\begin{eqnarray}
\frac{\D}{\D t} E_{\text{\tiny DP}}(t) = \gamma_{\text{\tiny DP}} - 
\xi_{\text{\tiny DP}} \,E_{\text{\tiny DP}}(t), \label{eq:eneq}
\end{eqnarray}
where
\begin{equation}
\gamma_{\text{\tiny DP}}=\frac{m\, G \,\hbar }{4\sqrt{\pi}(1+k)^3 R_0^3};\quad
\xi_{\text{\tiny DP}} = \frac{4 m^2\,G\, k}{3\sqrt{\pi} (1+k)^3 \,\hbar\, R_0}.
\end{equation}
So the mean value of the energy evolves in time as follows:
\begin{equation}\label{eq:enev}
 E_{\text{\tiny DP}}(t) =  E_{\text{\tiny DP}}(0)\,\, e^{-\xi_{\text{\tiny DP}} t} + \frac{\gamma_{\text{\tiny DP}}}{\xi_{\text{\tiny DP}}}
 \left(1-e^{-\xi _{\text{\tiny DP}}t}\right),
\end{equation}
which means that the energy relaxes exponentially to a finite value. 
By virtue of dissipation, now the system can loose energy as a consequence of the action of the collapse field and, even when the mean energy grows in time,
there is an upper value above which it cannot increase. 
Relying on the equipartition of energy, the finite asymptotic value of the energy can
be associated to the finite temperature of the collapse field:
\begin{eqnarray}\label{eq:temp}
T =\frac{2\gamma_{\text{\tiny DP}}}{3 k_B\,\xi_{\text{\tiny DP}}}=
\frac{\hbar^2}{8\, k_B}\,\frac{1}{m_r\, R_0^2 }.
\end{eqnarray}
Note that for $m_r = 0$ one finds an infinite temperature, which corresponds to the original DP model without dissipation.

It is worth noting how the temperature in Eq.\eqref{eq:temp}
does not depend on the mass of the system, which is a direct consequence of the coefficients
in front of the parameter $k$ in Eq.\eqref{eq:S0_Q}. In addition, one
can show how the master equation Eq.\eqref{eq:diss} with 
the operators 
$\hat{L}(\mathbf{\mom},\hat{\mathbf{p}})$
as in Eq.\eqref{eq:S0_Q} predicts 
a relaxation of the system's statistical operator to the stationary solution in the canonical form \cite{Vacchini2009}
\begin{equation} 
 \nu_{eq}(\hat{\bf p}) = (2 \pi M k_B T)^{-3/2} \exp\left(- \frac{\hat{\bf p}^2}{2 M k_B T}\right),
\end{equation} 
where $T$ is the temperature given by Eq.\eqref{eq:temp}.
As already remarked, the choice of the operators 
$\hat{L}(\mathbf{\mom},\hat{\mathbf{p}})$
in Eq.\eqref{eq:S0_Q}
is in principle not unique. Nevertheless, 
it directly follows  from the natural requirements 
that the dissipative DP model reduces to the original DP model for $k\rightarrow0$,
and that
it predicts a relaxation to the canonical stationary solution,
with a temperature $T$ which does not depend on the mass of the system.
Still, there are other possible choices which meet
these criteria, such as that provided by the operators
\begin{equation}
\hat{\tilde{L}}(\mathbf{\mom},\mathbf{\hat{p}}) \equiv  
 \frac{m\sqrt{G}}{\pi\sqrt{2}\,\hbar} 
 \,\frac{1}{\mom} \,
 \exp\left[-\frac{R^2_0}{2\hbar^2}
\left((1+k)\, {\mathbf{\mom}}+ 2k\, \hat{\mathbf{p}}\cdot{\bf \mom}\right)^2\right].
\end{equation}
However, one can see how a dissipative model with such operators would
lead to the same temperature as that in Eq.\eqref{eq:temp},
so that the conclusions of the present work would not be modified.

\section{Specification of Free Parameters in the Dissipative DP model}
The DP model contains one free parameter, which is $R_0$, the cut-off to regularize the DP dynamical equation; 
on the contrary, the dissipative DP model contains two free parameters: $m_r$, which is related with the temperature of the collapse field, 
and $R_0$. 
Anyway, contrary to $R_0$, the new parameter $m_r$ has a fully natural physical justification:
it controls the temperature of the collapse field.
The possibility of controlling the energy increase by controlling the temperature of the noise
opens the possibility of removing the arbitrary cut off $R_0 \sim 10^{-7} \text{m}$,
and replace it with the original cut off $R_0 \sim 10^{-15} \text{m}$ originally proposed 
by Di\'osi, corresponding to the nucleon's Compton wavelength and, as previously shown, to
the limit of validity of the non-relativistic approach.  In this section, we study
to what extent this natural
resolution of the overheating problem can be applied, by investigating
the possible choices of the parameters involved in the model.

For the value of temperature, one finds from Eq.\eqref{eq:temp}:
\begin{eqnarray}
T =\frac{10^{-19}}{m_r\,R_0^2}\,, \qquad\text{with the dimensions }\,[T]=\text{Kelvin}
\text{, }[m_r]=\text{amu} \text{, and }[R_0]=\text{m}.
\end{eqnarray}
If the collapse noise is real, then, whatever its nature, 
it must be a low-temperature noise (like the cosmic microwave background), with a temperature of few Kelvins; any other
choice would be difficult to justify. Accordingly, we have the following relation between the parameters of the dissipative DP model:
\begin{eqnarray}\label{eq:t1k}
T \sim 1 \,\text{K} \quad \Longrightarrow \quad m_r\, \,R_0^2\sim 10^{-19}, \qquad\text{with }\,[m_r]=\text{amu} 
 \text{, and }[R_0]=\text{m}.
\end{eqnarray}
By setting $R_0=10^{-15}\,$m in Eq.\eqref{eq:t1k}, one finds:
$m_r \sim 10^{11}$. But then,
for masses $m\ll m_r$, the dissipation parameter $k$ is very large, meaning a dissipation mechanism
which is too strong, as it involves an unrealistic amount of exchanged energy between the system
and the noise. 
In fact, for $k \gg 1$ the action of the noise would likely determine a sudden flip of the 
system's momentum, as one can infer by looking 
at the probability distribution of the exchanged momentum $P(\mathbf{\mom})$.
This is obtained by the operator $\hat{L}^2(\mathbf{\mom},\hat{\mathbf{p}})$ and hence
also depends on the momentum of the system.
Explicitly, for $k \gg1$, we have that if the system has momentum $\mathbf{p}$
the probability distribution associated with the exchanged momentum  is well approximated by:
\begin{equation}\label{eq:ll}
P(\mathbf{\mom}) =  \frac{C}{\mom^2} \exp\left(
-\frac{k^2 R^2_0}{\hbar^2}({\mom}+ 2 \mathbf{p}\cdot{{\bf \mom}}/\mom)^2\right),
\end{equation}
where $C$ is a normalization constant.
Hence, the most probable event is a decrease of momentum $2 p$ along the direction fixed by ${\bf p}$;
moreover, the width of the probability distribution is the narrower the higher the value of $k$. 
The momentum flip induced by the universal noise field causing the collapse of the wavefunction 
would mean that the latter can transfer an energy of the order
of tens of MeV to a nucleon in a nucleus
(corresponding to the average kinetic energy of a nucleon in a Fermi-gas model \cite{Bert}), which would induce instantaneous matter dissociation.
Accordingly, with $R_0=10^{-15}\,$m, the DP model is only an effective model for the mass ranges larger or comparable with $m_r\sim10^{11}\,$amu.
In other words, the DP model should be applied only to mesoscopic and macroscopic systems. 
A similar conclusion has been obtained by Di\'osi~\cite{d2}, from a different perspective.
It is worth mentioning that the value $m_r$ is also very different from the Plank mass $m_P\sim10^{19}\,$amu, which is sometimes considered as a borderline between quantum and classical masses~\cite{qg}.

For Ghirardi's cut-off value $R_0=r_c=10^{-7}\,$m, one finds: $m_r \sim 10^{-5}\,$amu. 
Relying on the analogy with collisional decoherence, see Eq.\eqref{eq:aT}, the dissipative effects are 
now much weaker compared with the case of the original cut-off since the mass of the environmental
particles is much smaller.

We can conclude that, unless one admits a limited applicability of the DP model, the
introduction of a cut-off of the order of $r_c$ cannot be avoided, even with the introduction of dissipation.

\section{Conclusion}
The goal of a gravity-induced collapse model is to explore the possibility ---perhaps only at the phenomenological level---that the collapse of the wave function is caused by gravity. Apart from much talking, in the literature only two such models (defined in terms of dynamical equations) have been proposed: the Schr\"odinger-Newton equation, and the Di\'osi-Penrose (DP) model. In~\cite{sneq} we have analysed the Schr\"odinger-Newton equation, and we have shown that in its present form it is not able to explain the collapse of the wave function in any satisfactory way.

In this paper, we have analysed the DP model. This model has the virtue of giving a dynamical explanation for the collapse rate suggested by Penrose. However, it does not succeed in a more ambitious goal: it does not explain why the collapse should be related to gravity. In the DP model, the noise (whose origin is not explored) does not couple to matter like a gravitational noise field is expected to do. This is in contrast with the Schr\"odinger-Newton equation, where the Newtonian self-interaction of different parts of the wave function is manifest. Contrary to the Schr\"odinger-Newton case, the DP model is structurally equivalent to the CSL model, the only difference being a different choice of the spatial correlation function of the noise. Therefore, from the conceptual point of view the DP model is on the same level as the GRW and CSL models, and it is not derived from more fundamental principles.

The original DP model was defined in terms of only the gravitational constant $G$, and no other parameter. However, because of divergences, one still needs to introduce a cut-off. A natural cut-off would correspond to the Compton wavelength of the nucleon, that justifies the non-relativistic approach. However, the model remains unrealistic, as it predicts a too high energy increase. A much higher---and less justified---cut-off is needed, to restore compatibility with known experimental facts.

In this paper, we explored a very natural way of dealing with the energy increase: we included dissipative terms in the dynamics,
which already allowed us to solve the problem of the energy divergence in both the GRW and the CSL model \cite{grwd,csld}. 
The motivation for such an approach is very simple: 
non-dissipative models (corresponding to an infinite-temperature bath/noise) are only idealization of more realistic situations, where the bath/noise has a finite temperature and the system interacting with it eventually thermalizes to that temperature. However, even in this case, the DP model remains unphysical, unless the cut-off is kept artificially large.

The conclusions seems to be that the DP model can be used to describe the collapse of massive composite systems, while its application to smaller systems such as atoms and nucleons is problematic.\\

\section*{Acknowledgements}
The authors acknowledge financial support from the EU project NANOQUESTFIT, INFN, and
the COST Action MP1006. The authors thanks L. Di\'osi, for his valuable comments on an early version of this paper.


\section*{References}

\begin{thebibliography}{1}

\bibitem{new_phys0} 
A. J. Leggett,  
{ J. Phys.: Condens. Matter} \textbf{14}, R415 (2002).

\bibitem{new_phys2} 
 S. L. Adler and A. Bassi, 
 { Science} {\bf 325}, 275 (2009).

\bibitem{new_phys1} 
S. L. Adler,  {\it Quantum Theory as an Emergent Phenomenon}, Cambridge University Press (2004).

\bibitem{new_phys3} 
S. Weinberg,  
{Phys. Rev. A} \textbf{85}, 062116 (2012).

\bibitem{Pe1}
P. Pearle, 
{ Phys. Rev. D} {\bf 13}, 857 (1976). 

\bibitem{GRW}
G.C. Ghirardi,  A. Rimini, and T. Weber, Phys. Rev. D {\bf{34}}, (470) (1986).

\bibitem{d}
L. Di\'{o}si, 
{ Phys. Lett. A} {\bf 120}, 377 (1987).

\bibitem{dd2}
L. Di\'{o}si,
{ Phys. Rev. A} {\bf 40}, 1165 (1989).

\bibitem{Pe2}
P. Pearle, 
{Phys. Rev. A} {\bf 39}, 2277 (1989).

\bibitem{Ghirardi1990b}
G.C. Ghirardi, P. Pearle, and A. Rimini, Phys. Rev. A {\bf{42}}, 78 (1990)

\bibitem{Cslmass}
G.C. Ghirardi, R. Grassi, F. and Benatti, 
{Found. Phys.} {\bf 25}, 5-38 (1995).

\bibitem{p}
P. Penrose, 
{ Gen. Rel. Grav.} {\bf 28}, 581 (1996).

\bibitem{adlerphoto}
S.L.Adler, 
{J. Phys. A} {\bf 40}, 2935 (2007).

\bibitem{Bassi2003}
A. Bassi and G.C. Ghirardi, Phys. Rep. {\bf{379}}, 257 (2003).

\bibitem{a}
S. L. Adler, 
{ J. Phys. A} {\bf 40}, 744 (2007).

\bibitem{d1}
L. Di\'{o}si, 
{ J. Phys. A: Math. Theor.} {\bf 40}, 2989 (2007);

\bibitem{Bassi2013}
A. Bassi, K. Lochan, S. Satin, T.P. Singh, and H. Ulbricht, Rev. Mod. Phys. {\bf{85}}, 471 (2013).


\bibitem{gravity_models1}
P. Pearle and E. Squires, Found. Phys. {\bf 26}, 291-305 (1996);

\bibitem{gravity_models2}
F. K\'arolyh\'azy, Nuovo Cimento A {\bf 42}, 390-402 (1966);
F. K\'arolyh\'azy, A. Frenkel, and B. Luk\'acs, in {\it Physics as Natural Philosophy}, A. Shimony and H. Feshbach, eds., MIT Press (1982);
F. K\'arolyh\'azy, A. Frenkel, and B. Luk\'acs, in
R. Penrose and C. J. Isham, eds., {\it Quantum Concepts in Space and Time}, Clarendon, Oxford (1986), pp. 109-146; A. Frenkel, Found. Phys. {\bf 20}, 159-188 (1990).

\bibitem{gravity_models3}
S. L. Adler, arXiv:1401.0353v3 (2014).

\bibitem{SN}
L. Di\'{o}si, 
Phys. Lett. A 105, 199 (1984); 
D. Giulini and A. Grossardt, Class. Quant. Grav. {\bf 29}, 215010 (2012).

\bibitem{Ghirardi1990}
G.C. Ghirardi, R. Grassi, and A. Rimini, Phys. Rev. A {\bf{42}}, 1057 (1990).

\bibitem{d2}
L. Di\'{o}si, arXiv:1404.6644, to appear in New J. Phys. (2014).

\bibitem{mw}
L. Mandel and E. Wolf, {\it Optical Coherence and Quantum Optics} (Cambridge
University Press, Cambridge, 1995), p.771.

\bibitem{Breuer2002}
H.-P. Breuer and F.~Petruccione, \emph{The Theory of Open Quantum Systems} (Oxford University Press, Oxford, 2002).

\bibitem{Lindblad1976}
G. Lindblad, Comm. Math. Phys. {\bf{48}}, 119 (1976);
V. Gorini, A. Kossakowski, and E.C.G. Sudarshan, J. Math. Phys. {\bf{17}}, 821 (1976).

\bibitem{Holevo1993}
A. S. Holevo, Rep. Math. Phys. {\bf 32}, 211 (1993).

\bibitem{gf}
M. R. Gallis G. N. and Fleming, 
{ Phys. Rev. A} {\bf 42}, 38 (1990).

\bibitem{a1}
S.L. Adler, 
{ J. Phys. A: Math. Gen.} {\bf 39}, 14067 (2006). 

\bibitem{Vacchini2009}
B. Vacchini and K. Hornberger, Phys. Rep. {\bf 478}, 71 (2009). 

\bibitem{Savage1985}
C. M. Savage and D. F. Walls, Phys. Rev. A {\bf{32}}, 2316 (1985).

\bibitem{Smirne2010}
A. Smirne and B. Vacchini, Phys. Rev. A {\bf{82}}, 042111 (2010).

\bibitem{Penrose1996}
R. Penrose, Gen. Relativ. Gravit. {\bf28}, 581 (1996).

\bibitem{Vacchini2007}
B. Vacchini, J. Phys. A: Math. Theor. {\bf{40}} 2463 (2007).

\bibitem{opto}
D. E. Chang {\it et al.}, PNAS {\bf 107}, 1005 (2010).

\bibitem{Bassi2005}
A. Bassi, E. Ippoliti, and B. Vacchini J. Phys. A: Math. Gen. {\bf{38}} 8017 (2005).

\bibitem{Vacchini2000}
B. Vacchini, Phys. Rev. Lett. {\bf{84}}, 1374 (2000).

\bibitem{gao}
S. Gao, Stud. Hist. Philos. Mod. Phys. 44, 148-151 (2013).

\bibitem{Bert}
C.A. Bertulani, P. Danielewicz {\it Introduction to Nuclear Reactions}
(IOP Publishing, London, 2004)

\bibitem{qg}
B. Lamine, R. Herv\'e, A. Lambrecht and S. Reynaud, Phys. Rev. Lett. {\bf 96}, 050405 (2006).

\bibitem{sneq}
M. Bahrami, A, Grossardt, and S. Donadi, A. Bassi,  New J. Phys. {\bf 16} 115007, (2014) .

\bibitem{grwd}
A. Smirne, B. Vacchini, and A. Bassi, Phys. Rev. A {\bf 90}, 062135 (2014).

\bibitem{csld}
A. Smirne and A. Bassi, arXiv:1408.6446, to appear in Sc. Rep. (2015).

\bibitem{d-private}
Private communication with L. Di\'osi.

\end{thebibliography}
\end{document}